\documentclass[a4paper]{article}
\usepackage[numbers]{natbib}

\usepackage{amsfonts,amssymb,amsmath}
\usepackage{epsfig,float}
\usepackage[ansinew]{inputenc}
\usepackage{tikz}

\usepackage{amsthm}

\usepackage{tikz}

\theoremstyle{plain}

\newtheorem*{theorem*}{Theorem}

\usepackage{multicol}

\usepackage{enumitem}

\usepackage{multirow}
\usepackage{setspace}
\usepackage{booktabs}
\usepackage{latexsym}
\usepackage{color}
\begin{document}
\begin{center}

\textbf{\large Robust and Sparse Multinomial Regression in High Dimensions}

\vskip 0.25cm

\textbf{Fatma Sevin\c{c} Kurnaz}$^{1}$\textbf{ and Peter Filzmoser}$^{2}$\\
{\small $^{1}$Yildiz Technical University, Istanbul, Turkey;
\textit{fskurnaz@yildiz.edu.tr} \\
$^{2}$Institute of Statistics and Mathematical Methods in Economics, \\TU Wien, Vienna, Austria;
\textit{peter.filzmoser@tuwien.ac.at}  }
\end{center}

\vskip 0.5cm {\centerline{\bf Abstract}}

\begin{quote}
\vskip -0.3cm 
A robust and sparse estimator for multinomial regression is proposed for high dimensional data. Robustness of the estimator is achieved by trimming the observations, and sparsity of the estimator is obtained by the elastic net penalty, which is a mixture of $L_1$ and $L_2$ penalties. From this point of view, the proposed estimator is an extension of the enet-LTS estimator \citep{Kurnaz18} for linear and logistic regression to the multinomial regression setting. After introducing an algorithm for its computation, a simulation study is conducted to show the performance in comparison to the non-robust version of the multinomial regression estimator. Some real data examples underline the usefulness of this robust estimator. 
\vskip 0.2cm

\par\end{quote}
\vskip 1cm


\section{Introduction} 

Multi-group classification is a widely discussed topic in statistics, and
there are various approaches. The most prominent method might be linear
discriminant analysis (LDA), where we typically have to assume that the
groups originate from multivariate normal distributions with equal
group covariances~\citep{JohW07}. Since the inverse pooled
covariance matrix is involved in the classification rule, 
such an approach would no longer work for high-dimensional
data with low sample size. This was solved by penalized  
versions of LDA, e.g.~\cite{hastie1995penalized}, and 
later also by versions, where the parameter estimates 
become sparse, thus contain many zeros in order to exclude
uninformative variables~\citep{clemmensen2011sparse}.

A difficulty which can challenge basically any data analysis
are outliers. In the context of classification, these can be ``unusual'' measurements being inconsistent with the observations of any of the groups, or observations with ``wrong'' group labels, which means that such an observation would have measurements that one would expect for another
group. Classification methods that are robust against outliers
have also been widely discussed in the literature; the
paper~\citep{Ortner20} discusses a robust approach for a 
sparse multi-group classification based on the optimal
scoring approach of~\cite{clemmensen2011sparse}.

Here we will focus on a different model for high-dimensional multi-group
classification, the multinomial regression model, which does
not require to specify the distribution of the data groups.
In the two-group case, this model reduces to logistic regression.
Logistic regression uses a vector of predictors
$\mathbf{x} \in \mathbb{R}^{p}$
to predict a binary response variable with group labels 
$G\in \{ 1,2 \}$.
If the number of groups is bigger than two, $K>2$,
the group labels are in the set $G=\{1,2,\dots,K \}$, and logistic regression extends to multinomial regression. 
In this case, the class conditional probabilities are modeled as
\begin{equation}
\label{eq:Mult}
Pr(G=l|\mathbf{x})=\frac{e^{\beta_{l0}+\mathbf{x}^T\pmb{\beta}_l}}{\sum_{j=1}^K e^{\beta_{j0}+\mathbf{x}^T\pmb{\beta}_j}}, \quad l=1,2,\ldots ,K ,
\end{equation}
with the intercept $\beta_{l0}$ and the parameter vector 
$\pmb{\beta}_l=(\beta_{l1},\ldots ,\beta_{lp})^T$ 
for modeling the outcome category $l$. 
Given a data matrix $\mathbf{X} \in \mathbb{R}^{n \times p}$ with rows
(observations) $\mathbf{x}_i \in \mathbb{R}^{p}$, for $i \in \{1,\ldots ,n\}$, 
and group labels $g_i \in G$,
the parameters are usually estimated by the maximum likelihood (ML) method.
For this we consider an indicator response matrix 
$\mathbf{Y} \in \mathbb{R}^{n \times K}$ with elements
$y_{il}=I(g_i=l)$, where $I(\cdot )$ denotes the indicator function.
The log-likelihood function is given by 
$\ell(\pmb{\beta})=\sum_{i=1}^n \sum_{l=1}^K y_{il} \log p_l(\mathbf{x}_i)$, where $p_l(\mathbf{x}_i)$ represents the predicted class probabilities, and $\pmb{\beta}$ denotes the matrix of dimension $(p+1)\times K$, with the parameters
$\beta_{l0}$ and $\pmb{\beta}_l$ in its columns.
Maximizing the log-likelihood function boils down to estimating the parameters 
in an iteratively reweighted least-squares scheme, which is no longer applicable
in a high dimensional setting, especially if $p\gg n$.
For this case it is common to use regularization, and in the context of 
multinomial regression different proposals exist \citep{Cawley06, Friedman10}.
Here we will focus on the elastic net estimator for multinomial regression introduced in \cite{Friedman10}, 
which is  based on a penalized form of the
log-likelihood function,
\begin{equation}
\label{eq:objMultenet}
\ell_{\lambda,\alpha}(\pmb{\beta})=
\frac{1}{n}\sum_{i=1}^n \left[ \sum_{l=1}^K  y_{il}
(\beta_{l0}+\mathbf{x}^T_i\pmb{\beta}_l)-
\log \left( \sum_{j=1}^K  e^{\beta_{j0}+\mathbf{x}^T_i\pmb{\beta}_j} \right) \right]  - \lambda P_{\alpha}(\pmb{\beta}) ,
\end{equation}
with the elastic net penalty
\begin{equation} 
\label{penalty}
P_{\alpha}(\pmb{\beta})= (1-\alpha)\frac{1}{2} \sum_{j=1}^p \sum_{l=1}^{K} \beta_{jl}^2  + \alpha \sum_{j=1}^p \sum_{l=1}^{K} |\beta_{jl}|  .
\end{equation}
The non-negative tuning parameter $\lambda$ controls the entire strength of the
penalty, while the tuning parameter $\alpha \in [0,1]$ allows to mix the
proportion of the ridge ($L_2$) and the lasso ($L_1$) penalty. This penalty structure provides to select variables like in lasso regression, and shrinks the coefficients according to ridge regression. An algorithm to estimate the unknown parameters has been
implemented in the R package \texttt{glmnet} \citep{Friedman21R}. 

A limitation of this estimator is its sensitivity with respect to outliers, which
goes back to the principle of ML estimation, where every observation contributes
equally to the penalized log-likelihood function. 
Different 
robust versions have been studied for the non-penalized
form of multinomial regression, \citep{Taba14, Yin18, Castilla18}, but generally they are not applicable in the high dimensional case. To the best of our knowledge,
there exists no robust version of multinomial regression for high-dimensional data.

The goal of this paper is to introduce a robust counterpart to the 
elastic net estimator for multinomial regression. 
The main idea is to use a trimmed version of the penalized log-likelihood function, similar as done in \cite{Kurnaz18} in the context of
sparse binary logistic regression.
The new estimator is introduced in detail in
Section~\ref{sec:motivation}. 
Section~\ref{sec:algorithm} provides an algorithm for the computation of the proposed estimator. The usefulness of the methodology is investigated in simulation studies in Section~\ref{sec:simulations}, and Section~\ref{sec:applications} shows the performance using real data examples. The final Section~\ref{sec:conclude} 
summarizes and concludes.


\section{A robust estimator for multinomial regression}
\label{sec:motivation}

The proposed robustified elastic net estimator for multinomial regression 
will be based on the idea of trimming the objective function 
given in Eq.~\eqref{eq:objMultenet}. 

\subsection{Definition of the estimator}

Every observation contributes to this objective
function, and large contributions can potentially originate from outliers, which
could be outliers in the explanatory variables, or a wrong class label, or both.
Thus, the idea is to just consider a subset of the observations to optimize the 
criterion. This is formulated in terms of the penalized negative trimmed
log-likelihood function as
\begin{equation}
\label{eq:objMultRob}
Q(H,\pmb{\beta}) =
- \frac{1}{h}\sum_{i \in H} \left[ \sum_{l=1}^K y_{il}(\beta_{l0}+\mathbf{x}^T_i\pmb{\beta}_l)-\log \left( \sum_{j=1}^K  e^{\beta_{j0}+\mathbf{x}^T_i\pmb{\beta}_j} \right) \right]  + h\lambda P_{\alpha}(\pmb{\beta}),
\end{equation} 
with the penalty of Eq.~(\ref{penalty}). Here, $H$ denotes a subset of the observations
of size $h$, thus $H \subseteq \{1,2,\dots,n\}$ and $\lvert H \rvert=h$.
The minimum of the objective function (\ref{eq:objMultRob}) determines the optimal subset of size $h$,
\begin{equation}
\label{eq:Hopt}
H_{opt} = \operatorname*{arg\,min}_{H \subseteq \{1,2,\dots,n\}:\lvert H \rvert=h} Q(H,\hat{\pmb{\beta}}_H) ,
\end{equation}
which is supposed to be outlier free.
The estimated coefficients $\hat{\pmb{\beta}}_H$ depend on the specific subset 
$H$, and result from the minimization problem
$\hat{\pmb{\beta}}_{H}=\operatorname*{arg\,min}_{\pmb{\beta}} Q(H,\pmb{\beta})$.
It is obvious that problem~(\ref{eq:Hopt}) would -- except for very small $n$ --
be computationally too expensive to be solved by considering all possible subsets of 
size $h$, and thus an approximate solution
has to be used. The resulting estimator will be denoted by
\begin{equation} 
\label{eq:robenet}
\hat{\pmb{\beta}}_{\mathrm{enetLTS}}=\operatorname*{arg\,min}_{\pmb{\beta}} Q(H_{opt},\pmb{\beta})
\end{equation}
and called enet-LTS estimator for multinomial regression.
The acronym LTS refers to the FAST-LTS algorithm, which goes back to
\cite{Rousseeuw06} in the robust regression setting.
This strategy to find an optimal subset has been employed for several robust
estimators, such as for linear and logistic regression with elastic net (enet)
penalty \citep{Kurnaz18}. The key feature of this algorithm are the 
C-steps (concentration steps), which works in the robust regression setting
as follows. Given an index set $H_m$ of size $h$ in the $m$-th
iteration of the algorithm. The regression parameters are estimated with these 
$h$ observations, and then residuals of all $n$ observations to the model can
be computed. The squared residuals are sorted, and the next subset $H_{m+1}$ of size $h$ 
in iteration $m+1$ is obtained by the indexes of the observations of the 
smallest $h$ squared residuals. It has been shown that the sum of squared residuals
for the subset $H_{m+1}$ is smaller or equal to that based on $H_m$. Thus, C-steps
are used to improve the value of the objective function.

\subsection{C-steps and robustness}
\label{sec:csteps}

In the context of multinomial regression, the criterion used within the C-steps 
is proposed as follows. Consider fixed parameters $\lambda$ and $\alpha$ for 
the penalty in Eq.~(\ref{eq:objMultRob}), and an $h$-subset $H_m$.
This will yield estimated coefficients $\hat{\pmb{\beta}}_{H_m}$. For simplicity we
denote the columns of this matrix by $(\hat{\beta}_{j0},\hat{\pmb{\beta}}_j^T)^T$, for $j=1,\ldots ,K$.
Then we can compute the ``scores'' or values of the linear link function
as $z_{i_{lj}}=\hat{\beta}_{j0}+\mathbf{x}^T_{i_l}\hat{\pmb{\beta}}_j$,
for $j=1,\ldots ,K$. The index $i_l$ refers to an observation of group $l$, where
$n_l$ is the number of observations in this group. The score vectors
$\mathbf{z}_{i_l}=(z_{i_{l1}},\ldots ,z_{i_{lK}})^T$ are then group-wise used for
multivariate outlier detection. The idea is that outlying scores in a group
could either indicate observations with a wrong group label, or observations
with very atypical $x$-values, or both. The next index set of the C-step,
$H_{m+1}$, will have to consist of the least outlying $h$ observations, where
the sizes of the groups have to be considered appropriately.

Group-wise multivariate outlier detection will be based on robust Mahalanobis distances
(RD) by using the Minimum Covariance Determinant (MCD) estimator \citep{Rousseeuw99}.
This requires a score matrix of full rank, but the matrix 
$\mathbf{Z}_l$ with rows
$\mathbf{z}_{i_l}$, for $i_l=1,\ldots ,n_l$, has at most rank $K-1$, because the 
probabilities in Eq.~(\ref{eq:Mult}) sum up to 1. Thus, we first use Singular
Value Decomposition (SVD) to decompose $\mathbf{Z}_l=\mathbf{U}_l \mathbf{D}_l
\mathbf{V}_l^T$, where the columns of these matrices are those corresponding to the 
$r_l$ non-zero singular values in $\mathbf{D}_l$. Denote the MCD estimates of location
and covariance of $\mathbf{U}_l$ by $\mathbf{t}_l$ and $\mathbf{C}_l$, respectively.
Then the robust Mahalanobis distances for the observations from group $l$ are
given as 
\begin{equation}
\label{eq:RD}
\mathrm{RD}(\mathbf{z}_{i_l})= \sqrt{(\mathbf{z}_{i_l}-\mathbf{t}_{l})^T\mathbf{C}_{l}^{-1}(\mathbf{z}_{i_l}-\mathbf{t}_{l})}, \quad i_l=1,\ldots ,n_l.
\end{equation}
In order to make these distances better comparable among the groups, we consider
group-wise scaled robust distances,
\begin{equation}
\label{eq:MultResRD}
\mathrm{RD}_s(\mathbf{z}_{i_l})=\mathrm{RD}(\mathbf{z}_{i_l})
\frac{\sqrt{\chi^2_{r_l,0.5}}}{\mathrm{median}_{i_l}\mathrm{RD}(\mathbf{z}_{i_l})}
\end{equation}
where $\chi^2_{r_l,0.5}$ is the $0.5$-quantile of the chi-square distribution with
$r_l$ degrees of freedom. Note that it would not be useful to just consider the
observations with the smallest scaled robust distances in the next iteration
for the C-step, because it could then happen that an entire group would be lost.
This can be avoided by using the same proportion of observations per group
as in the original sample also in the $h$-subsets.
Define $h_l=\lfloor (n_l+1)h/n \rfloor$ as the number of observations of group $l$
contained in a subset of size $h$, where the last number $h_K$ probably needs to
be adjusted such that $\sum_{l=1}^Kh_l=h$. The new index set $H_{m+1}$
in iteration $m+1$ of the C-steps will consist of those indexes corresponding to
the smallest $h_l$ scaled distances $\mathrm{RD}_s(\mathbf{z}_{i_l})$,
for $l=1,\ldots ,K$.

\subsection{Random starts with initial subsets}
\label{sec:randstart}

The global optimum is approximated by performing the C-steps with several random starts, based on so-called elemental subsets.
This idea has been outlined in \citep{Alfons13, Kurnaz18}, with the purpose
to keep the runtime of the algorithm low, but it needs to be adapted to multinomial regression.
For a certain combination of the penalty parameters $\alpha$ and $\lambda$, elemental subsets are created consisting of the indexes of two randomly selected observations from each category. Therefore, each elemental subset includes $2K$ randomly selected observations. We denote the $s$-th elemental subset by
\begin{equation}
\label{2obs}
H_{el}^s=\{\mathbf{j}_1^s,\mathbf{j}_2^s,...,\mathbf{j}_K^s\},
\end{equation}
where $\mathbf{j}_l^s$ refers to 2 randomly selected observation indexes 
$\{ j_{l1}, j_{l2} \}$ from category $l$, for $l=1,\ldots ,K$. In total we will
consider $s \in \{1,2,\dots,500 \}$ elemental subsets to compute the
estimator
\begin{equation} 
\label{eq:robenetH}
\hat{\pmb{\beta}}_{H_{el}^s}=\operatorname*{arg\,min}_{\pmb{\beta}} Q(H_{el}^s,\pmb{\beta}) ,
\end{equation}
where $Q(H_{el}^s,\pmb{\beta})$ is the objective function~(\ref{eq:objMultRob}), with $h$ replaced by $2K$.

Based on this estimator, the score matrix can be computed for all observations,
and scaled robust distances~(\ref{eq:MultResRD}) can be derived. 
Then two C-steps are carried out, starting with the $h$-subset identified by the indexes of the (group-wise) smallest scaled RD values.
This yields estimated parameters, say $\hat{\pmb{\beta}}$, which are used to compute the predicted
class probabilities
$\hat{p}_l(\mathbf{x}_i)=Pr(G=l|\mathbf{x}_i)$ for the $i$-th observation 
a particular class $l$,
see Eq.~(\ref{eq:Mult}).
The value of the objective function can then be denoted as
\begin{equation}
    \label{eq:Cstepobj}
    Q_{\lambda,\alpha}(\hat{H},\hat{\pmb{\beta}}) = -\frac{1}{h} \sum_{i=1}^h \sum_{l=1}^K {y_{il}\log \hat{p}_l(\mathbf{x}_i)}+h\lambda P_{\alpha}(\hat{\pmb{\beta}}),
\end{equation}
where $\hat{H}$ is the $h$-subset used for this estimator, and $P_{\alpha}(\hat{\pmb{\beta}})$ corresponds to the penalty term given 
in Eq.~(\ref{penalty}). 

Out of the 500 elemental subset starts, only those best $10$ $h$-subsets
with the smallest value of the objective function in~(\ref{eq:Cstepobj}) are kept.
With these $10$ candidate subsets, the C-steps are performed until convergence (no further decrease).
The result is an $h$-subset, called \textit{best subset}, which also
defines the estimator for this particular combination of $\alpha$ and $\lambda$.
The detailed algorithm in the next section will identify the optimal 
tuning parameters, and define the final estimator.

\section{Algorithm}
\label{sec:algorithm}

The steps outlined in the previous section can now be used to define the algorithm to compute the robustified elastic net estimator for multinomial regression. 

\medskip
\noindent
\textbf{Centering and scaling:}
At the beginning of the algorithm, the predictor variables are centered robustly by the median and scaled by the MAD. While carrying out the C-steps of the algorithm, we additionally center and scale the predictors by 
their arithmetic means and standard deviations, calculated on each current subset, see also \cite{Kurnaz18}. 
At the end, the coefficients are back transformed to the original scale.
 
\medskip
\noindent
\textbf{Tuning parameters:}
In order to identify the optimal tuning parameters $\alpha_{opt}$ and $\lambda_{opt}$, these parameters are varied on a specified grid, and the results need to be evaluated. In our experiments we have used 41 equally spaced points in $[0,1]$ for $\alpha$, and values from 0.05 to 0.95 in steps of 0.05 for $\lambda$ (with possible adjustment if the evaluation did not reveal a clear optimum). The procedure outlined in Section~\ref{sec:randstart} was carried out for the pair with the smallest $\alpha$ and $\lambda$ values. For further combinations we use the warm-start strategy as described in \citep{Alfons13,Kurnaz18}. That means we do not search an $h$-subset based on elemental subsets, but directly take the best $h$-subset from the neighboring grid value of $\alpha$ and/or $\lambda$, and start to perform the C-steps from this subset until convergence. Therefore, we obtain the best $h$-subsets for each combination of grid values $\alpha$ and $\lambda$.

\medskip
\noindent
\textbf{Evaluation using cross-validation (CV):}
Every combination of selected tuning parameters $\alpha$ and $\lambda$
results in a best subset containing $h$ observation indexes.
In order to determine the optimal combination of the tuning parameters $\alpha$ and $\lambda$ on the grid values, we are using $k$-fold CV as in \cite{Kurnaz18} (here we take $k=5$) to randomly split each of the
index sets into $k$ blocks of approximately equal size.
Already the $h$-subset consists of approximately the same proportions of
observations per group as the original data set, and also each of the CV folds
is constructed to consist of approximately the same proportions. However, here it could 
happen that one group has a very high proportion of outliers, and this should
be taken into account by an appropriately robustified evaluation measure.

The model is fit to the observations of $k-1$ blocks, and this yields 
class predictions $\hat{p}_l(\mathbf{x}_i)$ for the observations of the left-out
block, for each class $l=1,\ldots ,K$. This is done consecutively for each block being once the test-set block. Thus, we can compute deviances
$d_i=-\sum_{l=1}^K {y_{il}\log \hat{p}_l(\mathbf{x}_i)}$ for all observations
$i=1,\ldots ,h$ of the $h$-subset. In order to protect against outlying deviances
as a result from possibly poor predictions due to a high outlier proportion
in a group, we do not consider the 10\% of the largest deviances per group,
and thus just compute the mean of the smallest 90\% of the deviances per group as the 
robust evaluation measure.
Then the optimal parameter pair $\alpha_{opt}$ and $\lambda_{opt}$ is that combination of
$\alpha$ and $\lambda$ values which results in the minimum of 
this robust evaluation measure. The corresponding $h$-subset is
called $H_{opt}$. With this subset we minimize the objective 
function~(\ref{eq:objMultRob}) to obtain the estimator 
$\hat{\pmb{\beta}}_{opt}$.

\medskip
\noindent
\textbf{Reweighting step:}
Trimming can cause low efficiency of the estimator, and therefore a
reweighting step is added to increase the efficiency as in \cite{RousseeuwL03}. In a reweighting step, it is common that the outliers to the current model are identified and downweighted according to specific weights. The proposed weighting scheme
is using the scaled robust distances from Eq.~(\ref{eq:MultResRD}),
computed from $\hat{\pmb{\beta}}_{opt}$,
as follows:
\begin{equation}
\label{eq:reweights}
w_{i_l}=\begin{cases}
\quad 0,               & \mbox{ if } \mathrm{RD}_s(\mathbf{z}_{i_l})>c_2 \\
\quad 1,               & \quad \mathrm{else}\\
\end{cases} 
 \quad \quad i_l=1,\dots,n_l, \quad l=1,\ldots ,K,
\end{equation}
where $c_2=5$ is constant.
The weights $w_{i_l}$ can be mapped to weights $w_i$, for every observation $i=1,\ldots , n$.
The reweighted enet-LTS estimator is defined as 
\begin{equation}
\label{eq:rewest}
\hat{\pmb{\beta}}_{reweighted} = \operatorname*{arg\,min}_{\pmb{\beta}} \left\{ \sum_{i=1}^n w_i\left[ \sum_{l=1}^K y_{il}(\beta_{l0}+\mathbf{x}^T_i\pmb{\beta}_l)-\log \left( \sum_{j=1}^K  e^{\beta_{j0}+\mathbf{x}^T_i\pmb{\beta}_j} \right) \right]  + \lambda_{upd} n_w P_{\alpha_{opt}}(\pmb{\beta}) \right\},
\end{equation}
where $n_w$ is the number of nonzero weights. 
Since $h \leq n_w$, and because the optimal parameters $\alpha_{opt}$ and $\lambda_{opt}$ have been derived from $h$ observations, the penalty can act slightly differently in Eq. (\ref{eq:rewest}) than for the raw estimator. For this reason, the parameter $\lambda_{opt}$ has to be updated, while the $\alpha_{opt}$ compromising the tradeoff between the $L_1$ and the $L_2$ penalty is kept the same. The updated parameter $\lambda_{upd}$ is determined by $5$-fold CV and the $\alpha_{opt}$ is already fixed.

\section{Simulation studies}
\label{sec:simulations}

In this section, the enet-LTS estimator is compared to the classical non-robust 
multinomial logistic regression estimator with elastic net penalty 
\cite{Friedman10} by means of simulation studies. 
For the enet-LTS estimator we determine the subset size by
$h=\lfloor (n+1)\cdot 0.75\rfloor$. All simulations are carried
out in the statistical software environment R \citep{R21}.

To determine the optimal tuning parameters, we use the same 
procedure for the classical and robust estimator for coherence,
which means we first choose the same grid for $\alpha$ with 41 equally 
spaced points in $[0,1]$, and the same grid for $\lambda$ with values between 0.05 and 0.95 in 
steps of 0.05. 
Within a $5$-fold CV procedure, the mean of the deviances
is computed for the classical estimator, and the trimmed
mean for the robust estimator, see Sec.~\ref{sec:algorithm}. The minimum value
of this objective determines the optimal 
tuning parameters.

Note that we simulated the data sets with intercept. As described at the end of 
Section \ref{sec:motivation}, the data are centered and scaled at the beginning 
of the algorithm and only in the final step, the coefficients are back-transformed 
to the original scale, where the estimate of the intercept is computed.

\bigskip
\noindent
\textbf{Simulation schemes:}
We describe the general setting of the simulated data sets considering low-dimensional data
($p<n$) and high-dimensional data ($p>n$). Each data set consists of $K=3$ groups, each with $n_k$ observations (where
$n_1+n_2+n_3=n$), simulated from $p$-dimensional
normal distributions. 
In order to evaluate the effect on uninformative variables,
$p$ is split up into $p_a$ active variables which contribute to
the grouping information, and $p_b$ uncontributing noise
variables, thus $p=p_a+p_b$. Accordingly, the 
covariance matrix of the simulated data has a block structure: 
The elements of the covariance matrix of the first block of informative variables are
$\rho_a^{\mid j-k \mid}$, $1\leq j, k \leq p_a$,
and those for the second block are
$\rho_b^{\mid j-k \mid}$, $1\leq j, k \leq p_b$.
We will report results for $\rho_a=\rho_b=0.5$; those for 
other choices are qualitatively similar.
Both off-diagonal blocks of the covariance matrix have 
only zero entries.
The mean vectors of lengths $p$ of the groups are chosen as follows:
for group 1 we have $(3,3,0,\ldots ,0)^T$, for group 2 
we select $(3,-3,0,\ldots ,0)^T$, and for group 3 we have 
$(-3,-3,0,\ldots ,0)^T$. Thus, the groups are well separated
in two dimensions, but with increasing dimension (of the informative
variables) the separation gets more difficult.

The coefficient matrix $\pmb{\beta}$ consists of three columns for the three groups; the first row for the intercept terms is zero,
and also the block corresponding to the $p_b$ uninformative
variables is zero. For the block of the $p_a$ informative variables
we use the following entries: for the first group
the values $(0.5,0.5,\dots,0.5)^T$, for the second group
$(1,-1,1,-1,\ldots )^T$, and for the third group
$(-1,-1,\ldots ,-1)^T$.

For each observation, the category of the response variable $\mathbf{y}$ is randomly assigned according to the probabilities 
$Pr(G=l|\mathbf{x})=\frac{e^{\beta_{l0}+\mathbf{x}^T\pmb{\beta}_l}}{\sum_{j=1}^3 e^{\beta_{j0}+\mathbf{x}^T\pmb{\beta}_j}}$, for $l=1,2,3$.

Two different scenarios are considered for contamination. 
In the first scenario, outliers are added to the informative variables only. 
The second scenario includes outliers in both the informative and the uninformative variables. 
The outlier proportion $\varepsilon$ is set to 0\% (uncontaminated
data), 10\% and 20\%, respectively,
and outliers are generated by replacing the
first 10\% (or 20\%) of the observations of the corresponding
block of variables by random values independently
drawn from a normal distribution $N(10,1)$. 

We select the following five settings for the number of observations, and for the number of informative and uninformative variables, see Table~\ref{tab:settings}. The number of observations per group is (approximately)
equal.
Settings 1, 3 and 4 have $n>p$, and setting 2 and 5 have $p>n$.
Not only the proportion of $n$ to $p$ varies, but also the proportion of informative versus uninformative variables
varies quite a lot among the settings, see column 4.
The last three settings have equal sample size, and thus they
should reveal the effect of varying proportions of informative variables,
but also of increasing dimension.
\begin{table}[htbp]
\centering
\begin{tabular}{r|rrrrr}
Setting & $p_a$ & $p_b$ & $p_a:p_b$ & $p$ & $n$ \\
\hline
1 & 130 & 30 & $13:3$ & 160 & 500 \\
2 & 250 & 250 & $1:1$ & 500 & 300 \\
3 & 50 & 100 & $1:2$ & 150 & 180 \\
4 & 5 & 50 & $1:10$ & 55 & 180 \\
5 & 50 & 950 & $1:19$ & 1000 & 180
\end{tabular}
\caption{Different settings for the simulations.}
\label{tab:settings}
\end{table}

\bigskip
\noindent
\textbf{Performance measures:}
To evaluate the different estimators, training and test data sets are generated according to the sampling schemes 
explained earlier. The models are fit to the training data and evaluated on the test data. The test data are always 
generated without outliers.

As performance measures we use the misclassification rate (MCR) defined by
\begin{equation}
\mathrm{MCR}=\frac{m}{n}
\label{eq:mcr}
\end{equation}
where $m$ is the number of misclassified observations from the test data after fitting the model on the training data.

A further quality criterion is the precision of the coefficient estimator for the informative and uninformative variables: 
\begin{equation}
\label{eq:biasinf}
\mathrm{PRECISION_{inf}}(\hat{\pmb{\beta}})=\sqrt{\frac{1}{p_aK}\sum_{j=2}^{p_{a}+1}\sum_{l=1}^K \left(\beta_{jl}-\hat{\beta}_{jl} 
\right)^2},
\end{equation}
and
\begin{equation}
\label{eq:biasnoninf}
\mathrm{PRECISION_{uninf}}(\hat{\pmb{\beta}})=\sqrt{\frac{1}{p_bK}\sum_{j=p_{a}+2}^{p+1}\sum_{l=1}^K \left(\beta_{jl}-\hat{\beta}_{jl} 
\right)^2} .
\end{equation}

Concerning the sparsity of the coefficient estimators, we evaluate the False Positive Rate (FPR) and the False Negative Rate (FNR), defined as
\begin{equation}
\label{eq:fpr}
\mathrm{FPR}(\hat{\pmb{\beta}})=\frac{\lvert \{ j=1,\dots,p \quad \& \quad l=1,\dots,K:\hat{\beta}_{jl} \neq 0 \wedge \beta_{jl}=0 \} \rvert}{p_b \cdot K},
\end{equation}
\begin{equation}
\label{eq:fnr}
\mathrm{FNR}(\hat{\pmb{\beta}})=\frac{\lvert \{ j=1,\dots,p \quad \& \quad l=1,\dots,K:\hat{\beta}_{jl}=0 \wedge \beta_{jl} \neq 0 \} \rvert}{p_a \cdot K},
\end{equation}
respectively.
The FPR is the proportion of uninformative variables that are incorrectly included in the model. On the other hand, the FNR is the proportion of informative variables that are incorrectly excluded from the model. 
A high FPR usually has a bad effect on the prediction performance since it inflates the variance of the estimator.
Also high FNR is undesirable, since this can lead to wrong interpretations.

These evaluation measures are calculated for the generated data in each of 100 simulation replications separately. 
The evaluation measures are averaged over the replications and are summarized in the figures below. 
The smaller the value for these criteria, the better the performance of the method.

\begin{figure}[htbp]
\includegraphics[width=0.5\textwidth]{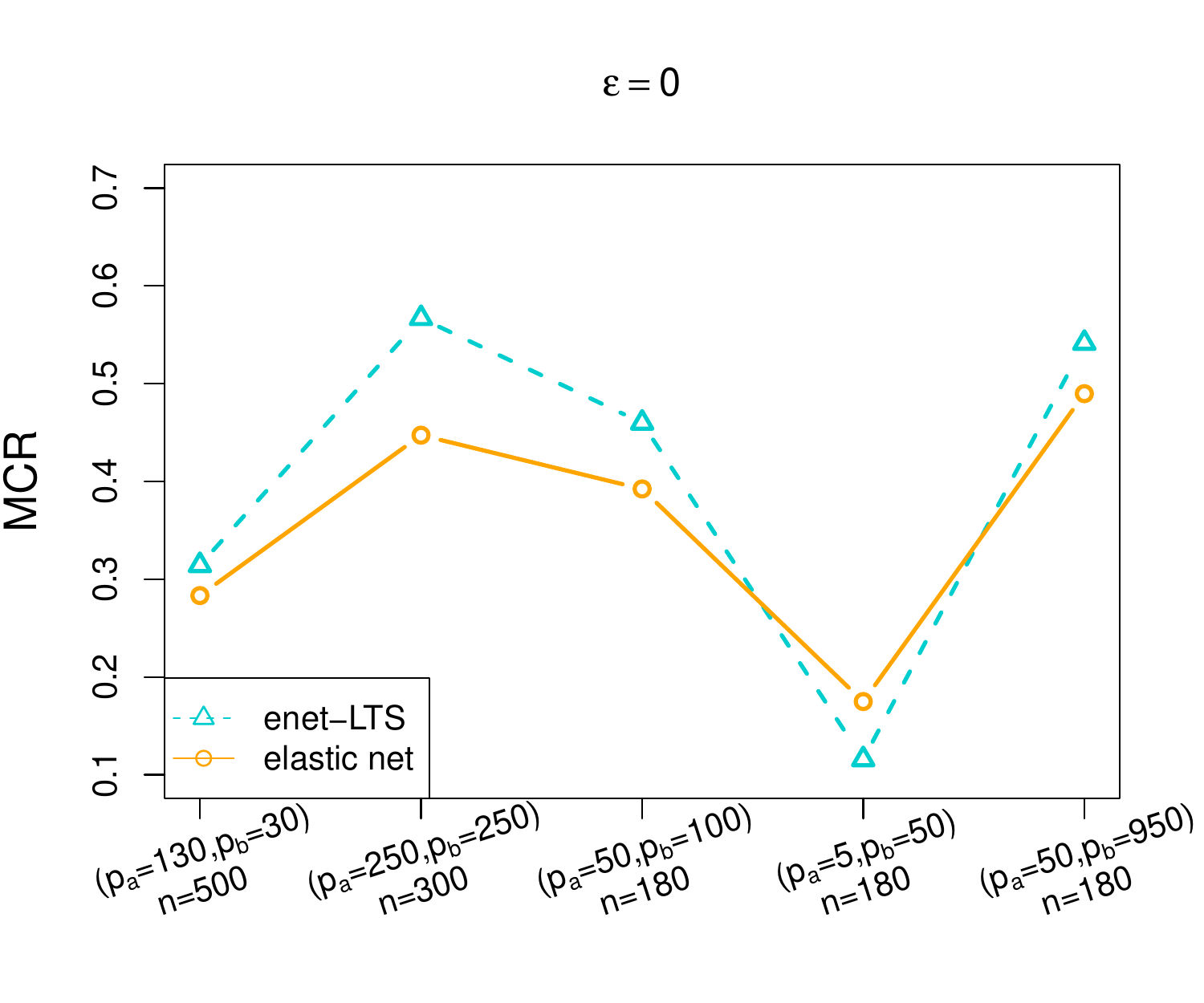}
\hfill
\includegraphics[width=0.5\textwidth]{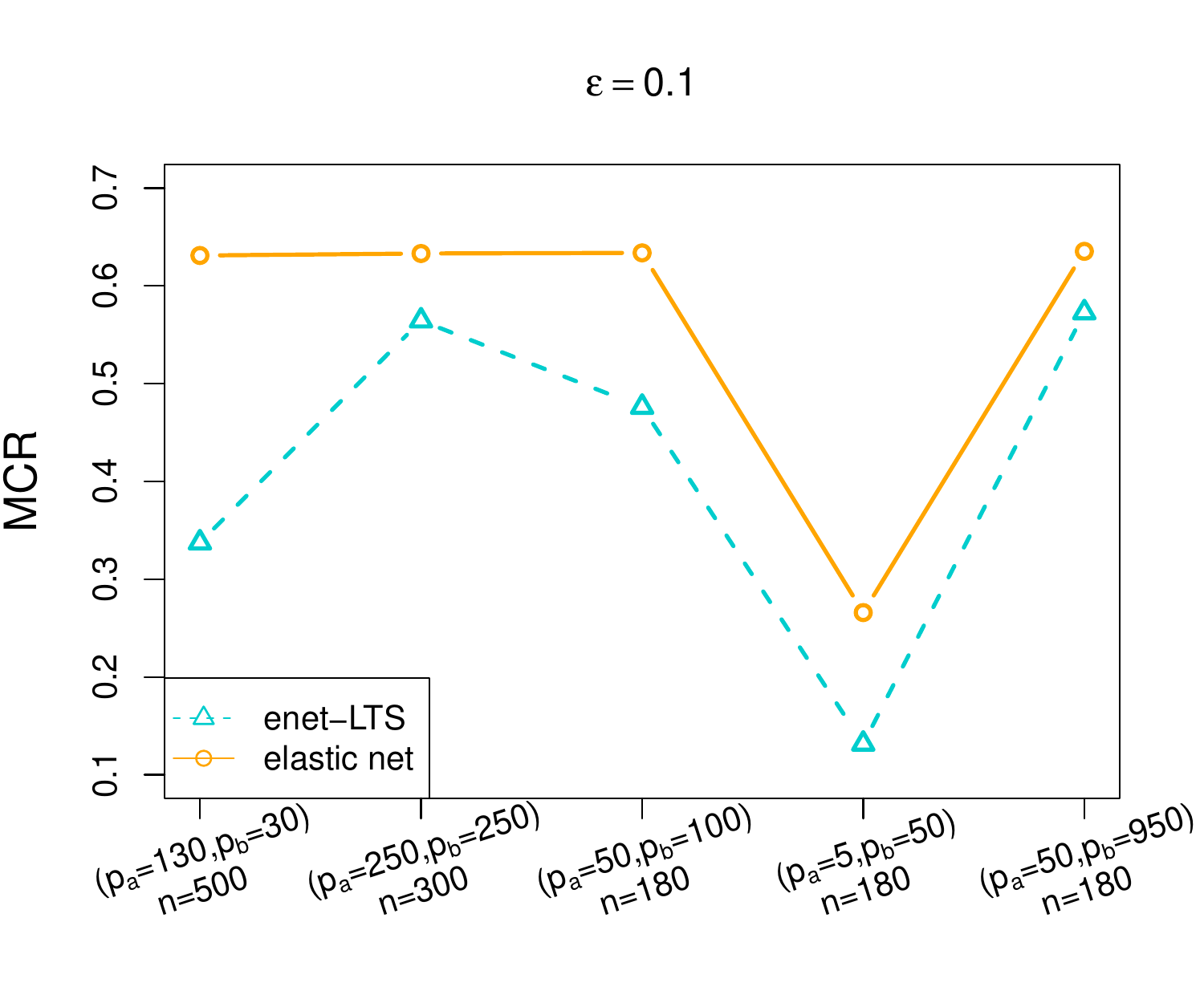}
\caption{Misclassification rate (MCR) for the different 
settings. Left: clean data; right: 10\% contamination in informative variables only.}
\label{fig:mcr}
\end{figure}

\begin{figure}[htbp]
\includegraphics[width=0.5\textwidth]{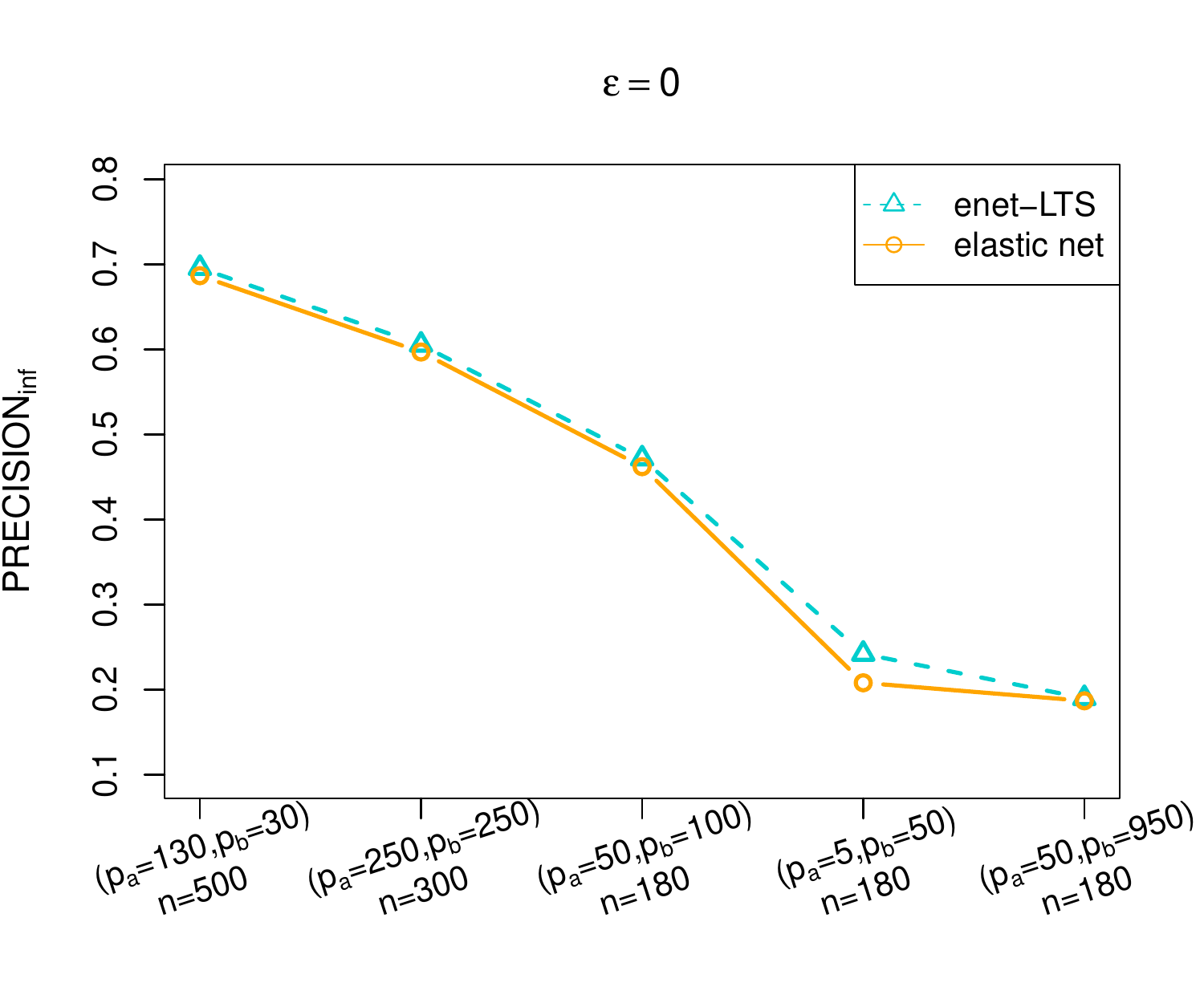}
\hfill
\includegraphics[width=0.5\textwidth]{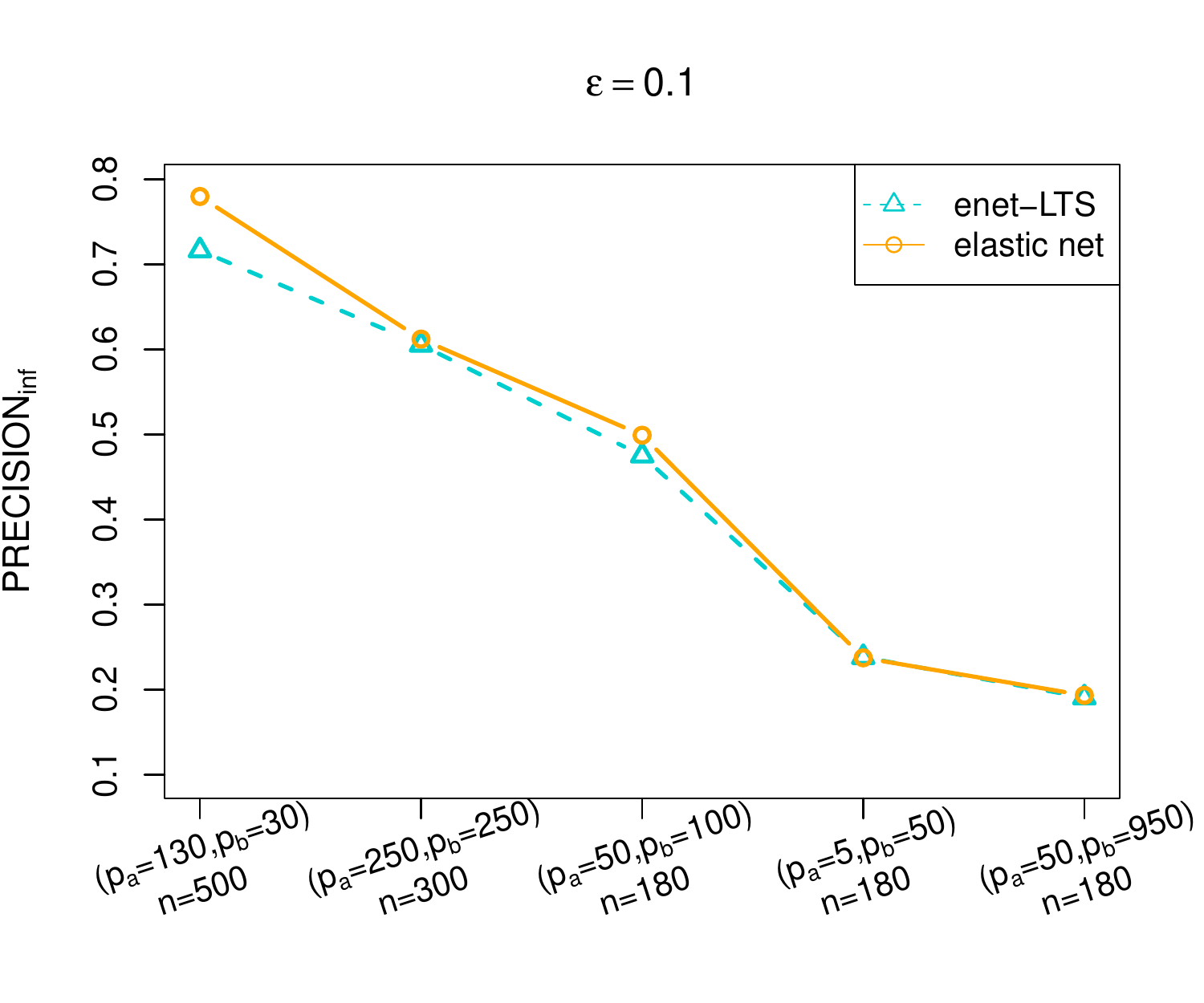}
\caption{Precision of the estimated coefficients for informative variables ($\mathrm{PRECISION_{inf}}$) for the different settings. Left: clean data; right: 10\% contamination in informative variables only.}
\label{fig:pinf}
\end{figure}

\begin{figure}[htbp]
\includegraphics[width=0.5\textwidth]{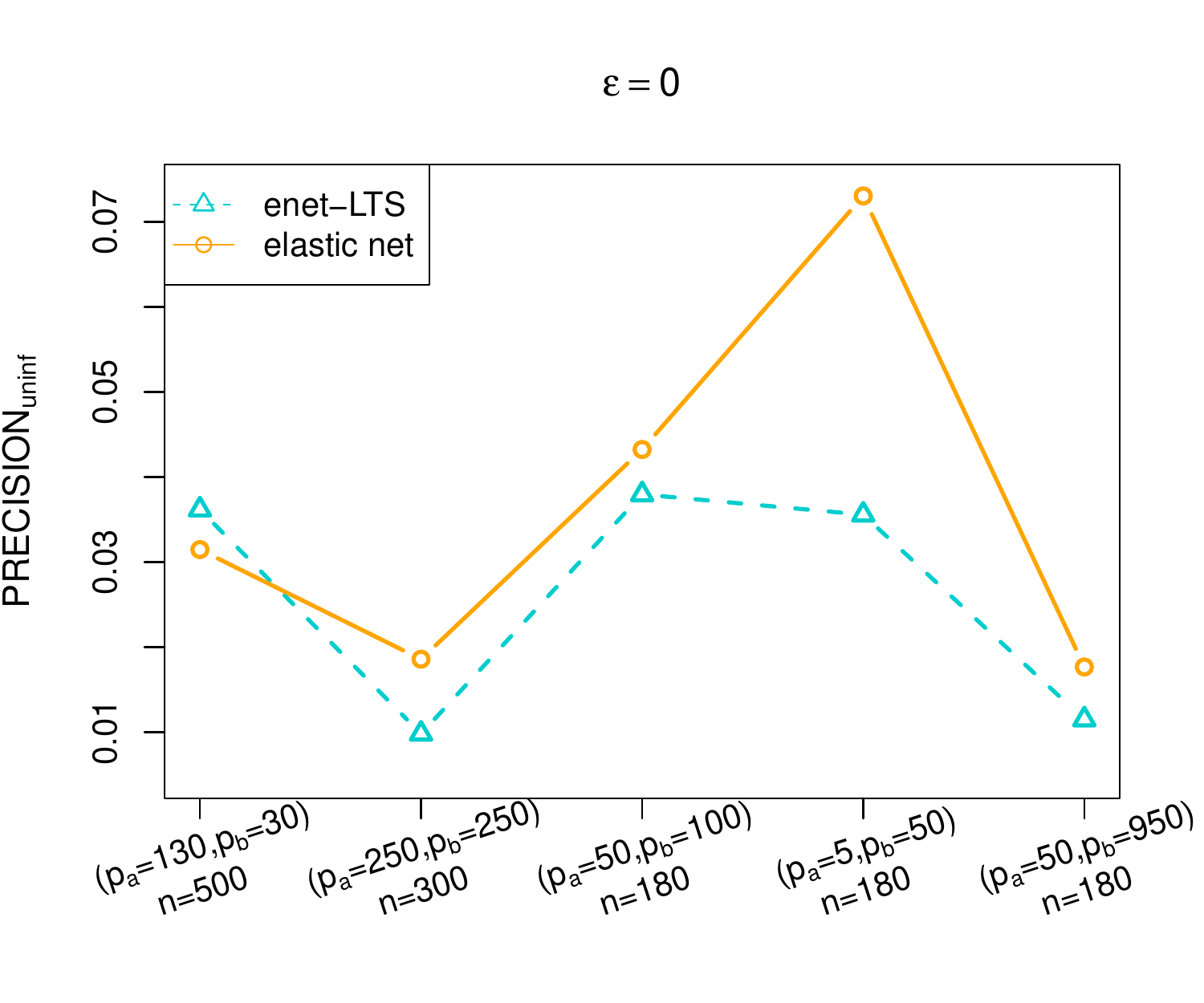}
\hfill
\includegraphics[width=0.5\textwidth]{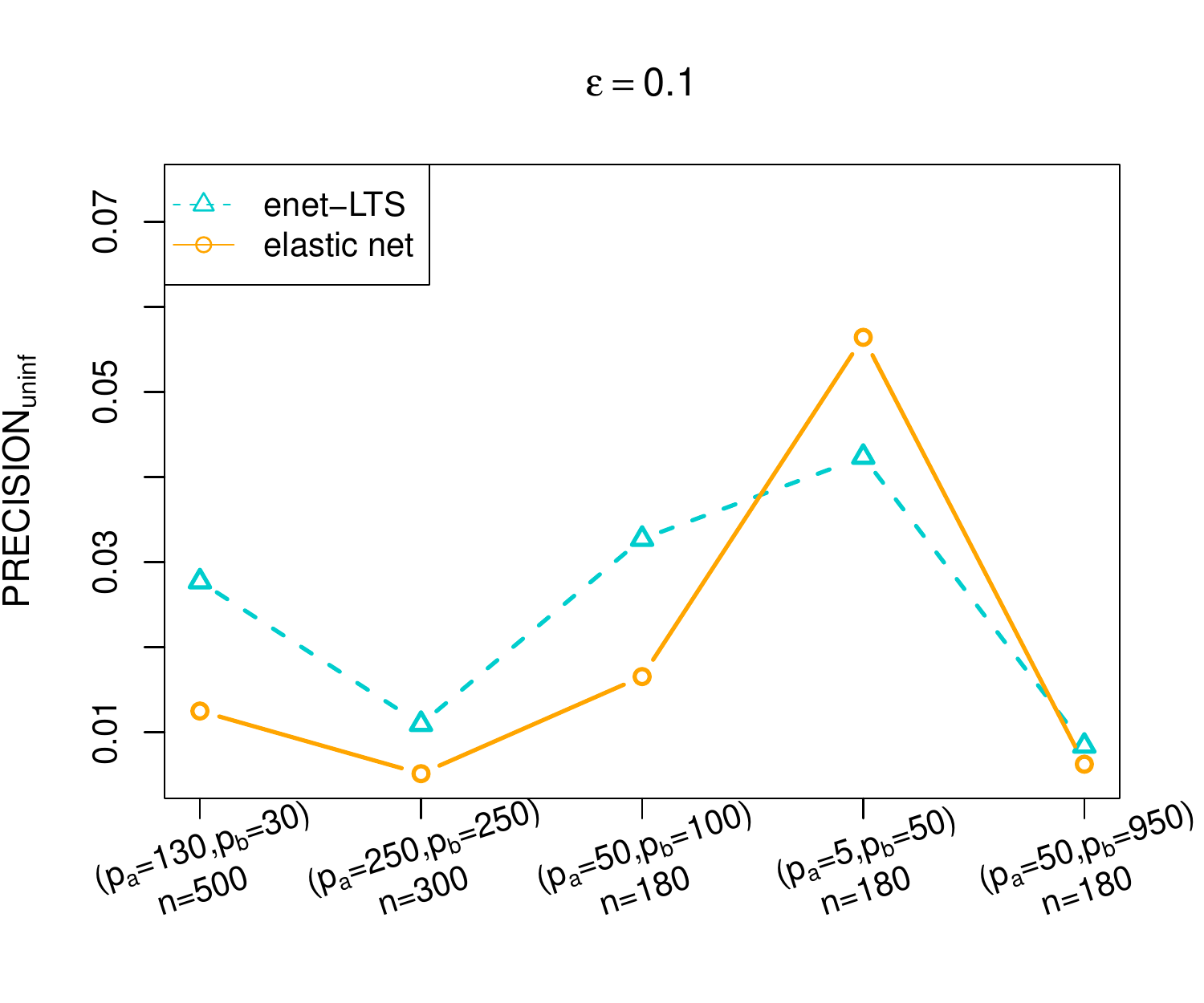}
\caption{Precision of the estimated coefficients for uninformative variables ($\mathrm{PRECISION_{uninf}}$) for the different settings. Left: clean data; right: 10\% contamination in informative variables only.}
\label{fig:pnoninf}
\end{figure}

\begin{figure}[htbp]
\includegraphics[width=0.5\textwidth]{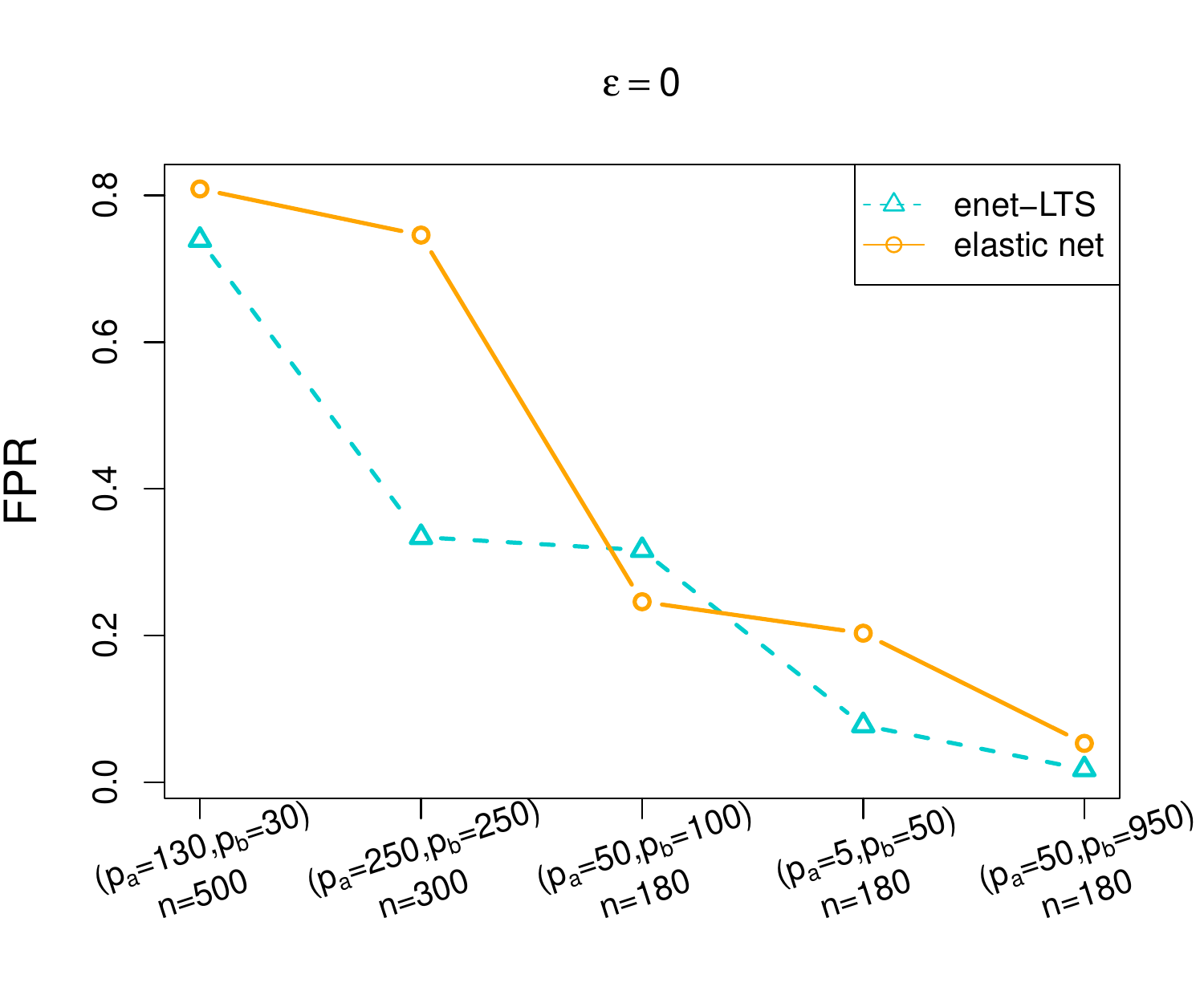}
\hfill
\includegraphics[width=0.5\textwidth]{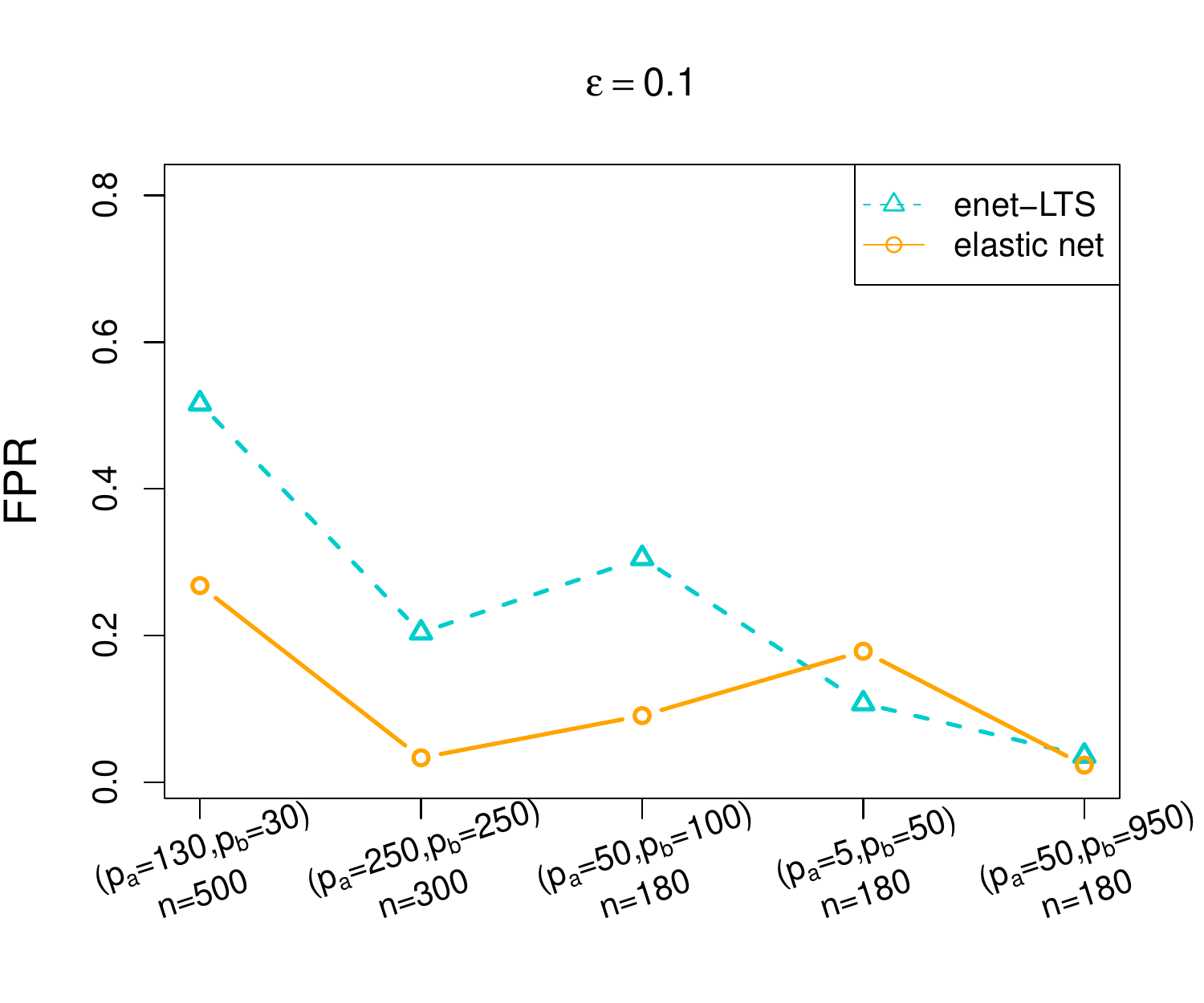}
\caption{False positive rate (FPR) for the different scenarios. Left: clean data; right: 10\% contamination in informative variables only.}
\label{fig:fpr}
\end{figure}

\begin{figure}[htbp]
\includegraphics[width=0.5\textwidth]{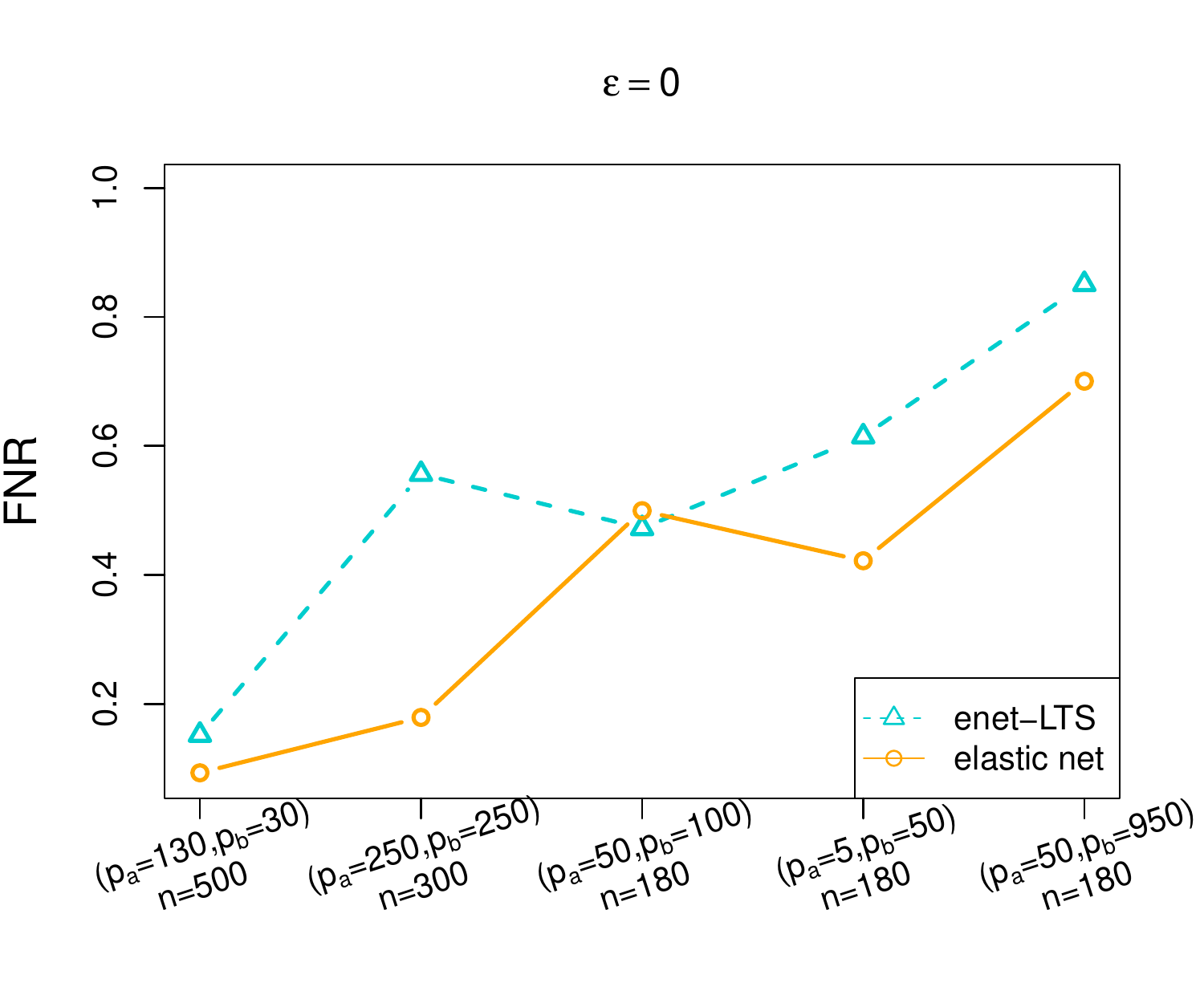}
\hfill
\includegraphics[width=0.5\textwidth]{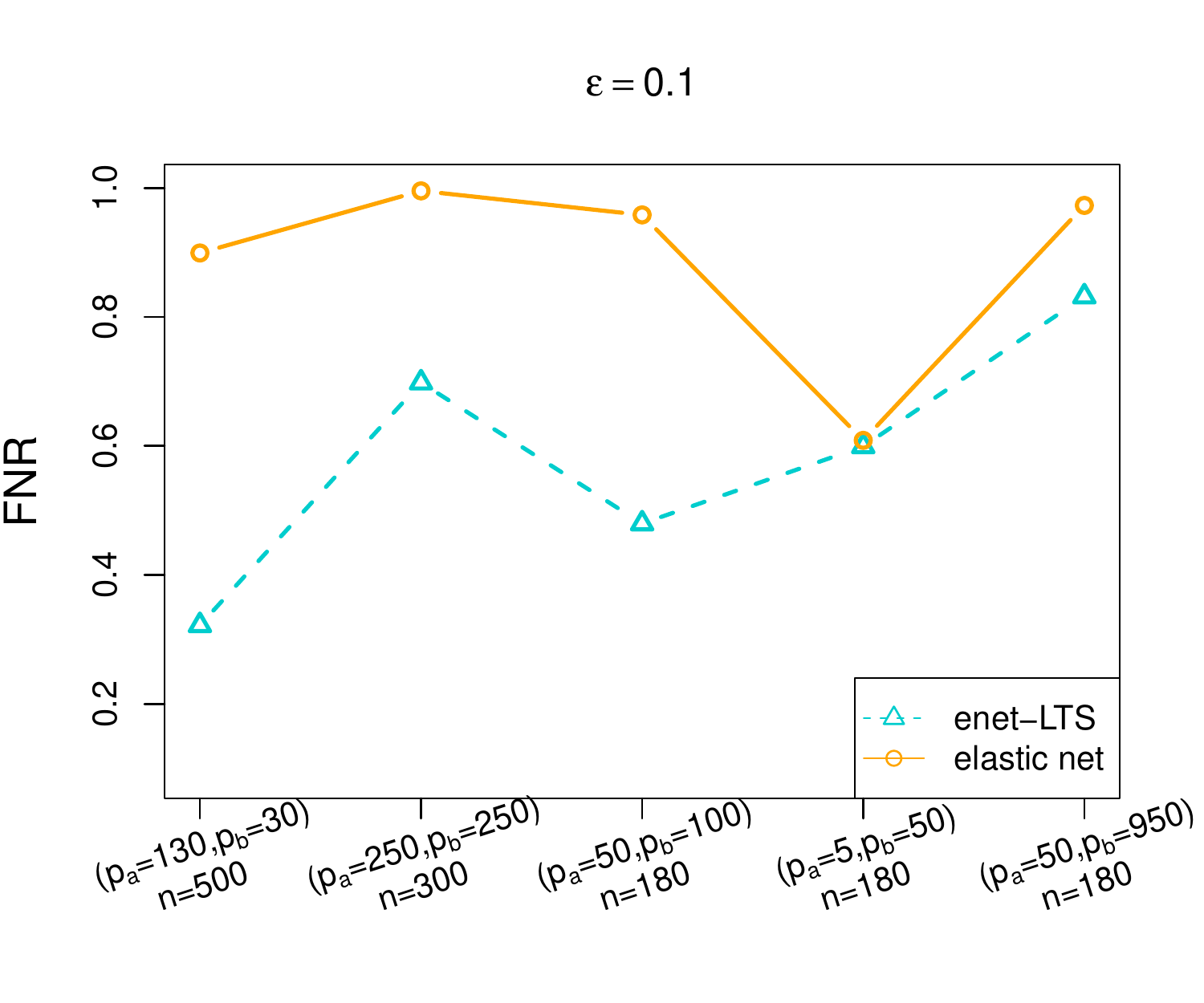}
\caption{False negative rate (FNR) for the different scenarios. Left: clean data; right: 10\% contamination in informative variables only.}
\label{fig:fnr}
\end{figure}

\bigskip
\noindent
\textbf{Results:}
The results of the simulations are presented in 
Figs.~\ref{fig:mcr}-\ref{fig:fnr}, and the horizontal axes are
arranged according to settings 1-5 (Table~\ref{tab:settings}),
thus sorted according to a decreasing ratio of informative versus 
uninformative variables. The left plots show the outcomes for
clean uncontaminated data, while the right plots are for 
results with 10\% contamination.
We also computed the results $20$\% contaminated, but since they
lead to the same conclusions as those for 10\% contamination,
they are omitted here.
Moreover, we only present the results for outliers in the 
informative variables, and omit those where outliers are in both
data parts, as they reflect a similar structure. 

Fig. \ref{fig:mcr} shows the misclassification rates for the
different settings. Note that according to the data generation, 
the group separation becomes more difficult in higher dimension.
This explains a relatively low MCR for setting~4 with $p=55$, and the 
highest MCR for setting 5 with $p=1000$. For uncontaminated data 
($\varepsilon=0$), the MCR of the elastic net estimator is smaller
than that for the enet-LTS, except for setting 4, which is a bit 
unexpected. In this setting, the precision for the uninformative
variables is much better for the robust than for the classical method
(Figure~\ref{fig:pnoninf} left), and also the FPR is smaller
(Figure~\ref{fig:fpr} left), which explains the discrepancy for
MCR. With 10\% contamination, the MCR increases in all settings for the 
non-robust method, while it is nearly unchanged for the robust one.

The precision of the informative variables in Fig.~\ref{fig:pinf} decreases with the decreasing rate of $p_a/p_b$, thus with increasing 
model sparsity. 
This decrease is also connected to a decrease of FPR
(Fig.~\ref{fig:fpr} left) and an increase in FNR
(Fig.~\ref{fig:fnr} left), for increasing sparsity.
This means that for less sparse models, the methods tend to include
noise variables, while for sparser models, they tend to exclude
informative variables. Thus, the block of non-zero regression coefficients might be easier to being recovered if sparsity increases,
and if estimated zeros appear in this block.

The precisions for the informative variables are slightly better
for the classical elastic net estimator in the uncontaminated 
case, and slightly better for enet-LTS with contaminated data 
(Fig.~\ref{fig:pinf}). 
The precision of the uninformative variables in Fig.~\ref{fig:pnoninf} 
is quite comparable for the two estimators in the uncontaminated case, except for setting 4. Interestingly, enet-LTS is a bit better for 
the uncontaminated data, but elastic net has better performance in
the contaminated case. However, when comparing the scale for the 
precision of the informative and uninformative variables, we see
that these differences in the uninformative case are quite 
marginal.

Fig. \ref{fig:fpr} and Fig. \ref{fig:fnr} show the quality of the 
variable selection ability of the estimators with respect to FPR
and FNR. Both measures complement each other: when FPR increases,
FNR decreases, and vice versa. The picture for enet-LTS is quite
comparable in the uncontaminated and contaminated case, while for 
the classical estimator we see a completely different behavior:
In case of contamination, the FNR increases enormously in almost all
settings, thus far too few informative variables are selected.
Compared to the uncontaminated case, the FPR can be reduced, which means
a reduction of incorrectly selected noise variables. In other
words, the classical estimator tends to produce much higher sparsity
in presence of contamination.

\clearpage
\section{Real data applications} 
\label{sec:applications}

\subsection{Analysis of handwritten digits}

We consider a data set with normalized handwritten digits, which is available
from the R package \citep{tensorBSS} and goes back to \cite{matan1990handwritten}. 
This quite famous data set has been used a lot in machine learning, and it 
consists of images with scanned digits from envelopes of delivered letters from the
the U.S. Postal Service. The images with the digits 0 to 9 (thus there are 10 groups)
have been preprocessed to 16 $\times$ 16 grayscale images.
We fit our model to the training set consisting of 7291 observations, and 
evaluate for the test set which has 2007 observations. 
Table~\ref{tab:digittrain} shows the resulting confusion table for the training
data, and Table~\ref{tab:digittest} that for the test data.
The correct classification rate is 88.1\% for the training data, and
84.3\% for the test data. One can see that the misclassifications are quite
different among the different digits, depending on whether these are very similar
or not. For example, digit ``4'' is frequently misclassified as digit ``1'',
``2'', or ``9''.
Compared to other classifiers, especially from machine learning, see \texttt{http://yann.lecun.com/exdb/mnist/}, our results for the correct classification rate are not too exciting, but we can get interesting interpretations, as seen in the following.

\begin{table}[htbp]
\centering
\begin{tabular}{rr|rrrrrrrrrr}
&   & \multicolumn{10}{c}{\textbf{Predicted class}} \\[2mm]
&  & 0 & 1 & 2 & 3 & 4 & 5 & 6 & 7 & 8 & 9 \\ 
  \hline
\multirow{10}{0.5cm}[-0.1cm]{\rotatebox{90}{\textbf{True class}}}
&  0 & 1067 &   0 &  20 &   7 &  15 &  10 &  41 &   6 &  23 &   5 \\ 
&  1 &   0 & 996 &   0 &   0 &   0 &   0 &   0 &   0 &   8 &   1 \\ 
&  2 &  13 &   1 & 640 &  13 &  14 &   0 &   3 &  14 &  27 &   6 \\ 
&  3 &   1 &   1 &  14 & 576 &   2 &  22 &   0 &  10 &  24 &   8 \\ 
&  4 &   1 &  34 &  25 &   0 & 551 &   1 &  11 &   0 &   9 &  20 \\ 
&  5 &  12 &   1 &  10 &  25 &  31 & 456 &   4 &   5 &   8 &   4 \\ 
&  6 &  10 &   4 &  39 &   0 &  13 &  16 & 568 &   0 &  14 &   0 \\ 
&  7 &   1 &   2 &  17 &   0 &  34 &  10 &   0 & 529 &   4 &  48 \\ 
&  8 &   7 &   3 &  10 &  19 &  14 &  11 &   3 &   3 & 465 &   7 \\ 
&  9 &   4 &   2 &   2 &   1 &  12 &  17 &   0 &  23 &   8 & 575 \\ 
   \hline
\end{tabular}
\caption{Confusion table for the training data of the handwritten digits data set.}
\label{tab:digittrain}
\end{table}

\begin{table}[htbp]
\centering
\begin{tabular}{rr|rrrrrrrrrr}
&   & \multicolumn{10}{c}{\textbf{Predicted class}} \\[2mm]
&  & 0 & 1 & 2 & 3 & 4 & 5 & 6 & 7 & 8 & 9 \\ 
  \hline
\multirow{10}{0.5cm}[-0.1cm]{\rotatebox{90}{\textbf{True class}}}
& 0 & 317 &   0 &  11 &   3 &   4 &   1 &  15 &   0 &   4 &   4 \\ 
& 1 &   0 & 251 &   1 &   2 &   4 &   0 &   2 &   0 &   2 &   2 \\ 
&  2 &   6 &   0 & 157 &   6 &   9 &   1 &   1 &   2 &  16 &   0 \\ 
&  3 &   5 &   0 &   4 & 128 &   3 &  14 &   0 &   1 &   6 &   5 \\ 
&  4 &   1 &   9 &   7 &   0 & 164 &   0 &   3 &   1 &   4 &  11 \\ 
&  5 &   8 &   0 &   0 &  10 &   9 & 123 &   0 &   0 &   4 &   6 \\ 
&  6 &   5 &   0 &  10 &   0 &   3 &   7 & 142 &   0 &   3 &   0 \\ 
&  7 &   1 &   2 &   4 &   0 &  15 &   0 &   0 & 118 &   1 &   6 \\ 
&  8 &   4 &   0 &   1 &  12 &   5 &   4 &   2 &   1 & 132 &   5 \\ 
&  9 &   0 &   2 &   0 &   1 &   4 &   2 &   0 &   7 &   2 & 159 \\ 
   \hline
\end{tabular}
\caption{Confusion table for the test data of the handwritten digits data set.}
\label{tab:digittest}
\end{table}

\begin{figure}[htp]
\includegraphics[width=\textwidth]{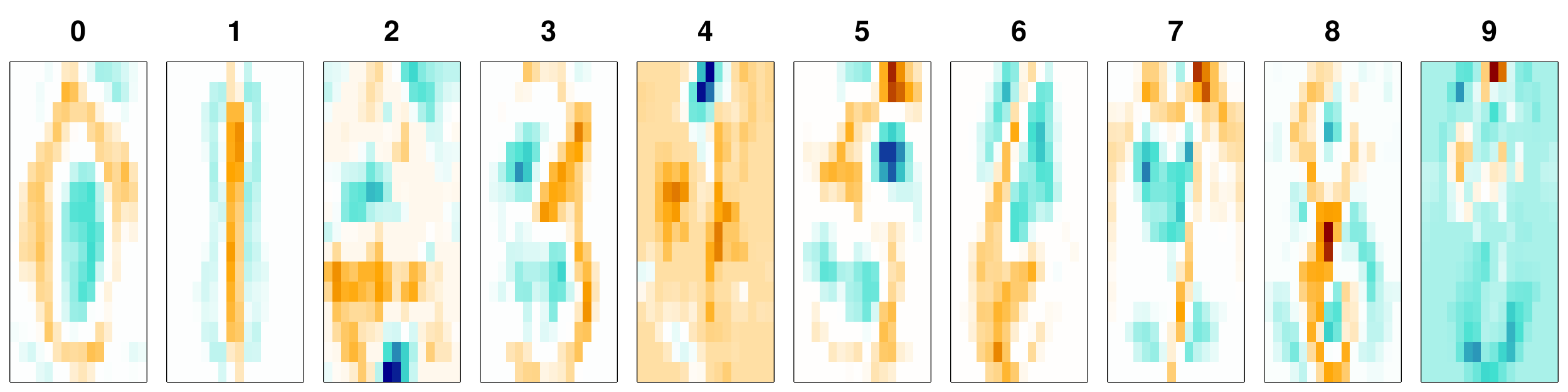}
\caption{Visualization of the estimated regression coefficients.}
\label{fig:coef}
\end{figure}

In order to understand how the classifier works and which pixel information is
important for the class assignments, we show a plot of the regression coefficient
matrix in Figure~\ref{fig:coef}. Every plot is for one digit (legend on top),
and consists of 16 $\times$ 16 regression coefficients presented as image.
Darker color means higher coefficient, red for positive, blue for negative.
White refers to zero coefficients. Some of the coefficients are quite sparse,
such as those for digit ``1'', others have almost no zeros. Since the input data
are scaled in $[-1,1]$, and higher values refer to darker image information,
we can say that positive coefficients increase the probability of assignment to 
a class. Accordingly, Figure~\ref{fig:coef} shows well with red color which
pixels are important for the class assignment, and which contribute 
negatively (blue). There seem to be some specifics related to the style
how the digits are written, seen for example in the upper right part of
digit ``5''.

\begin{figure}[htp]
\includegraphics[width=\textwidth]{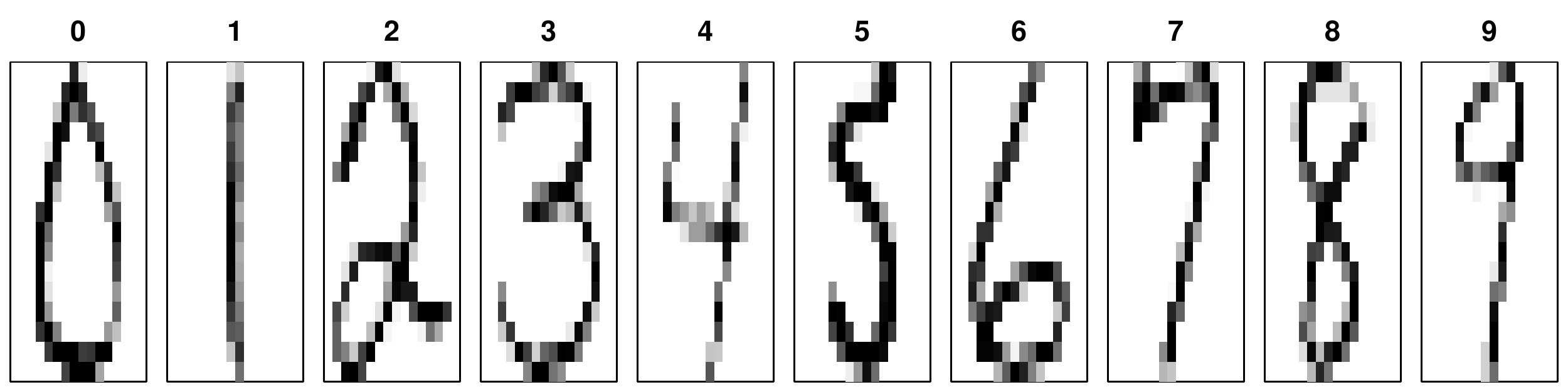}
\caption{Most ``inlying'' digits according to the model.}
\label{fig:digin}
\end{figure}

\begin{figure}[htp]
\includegraphics[width=\textwidth]{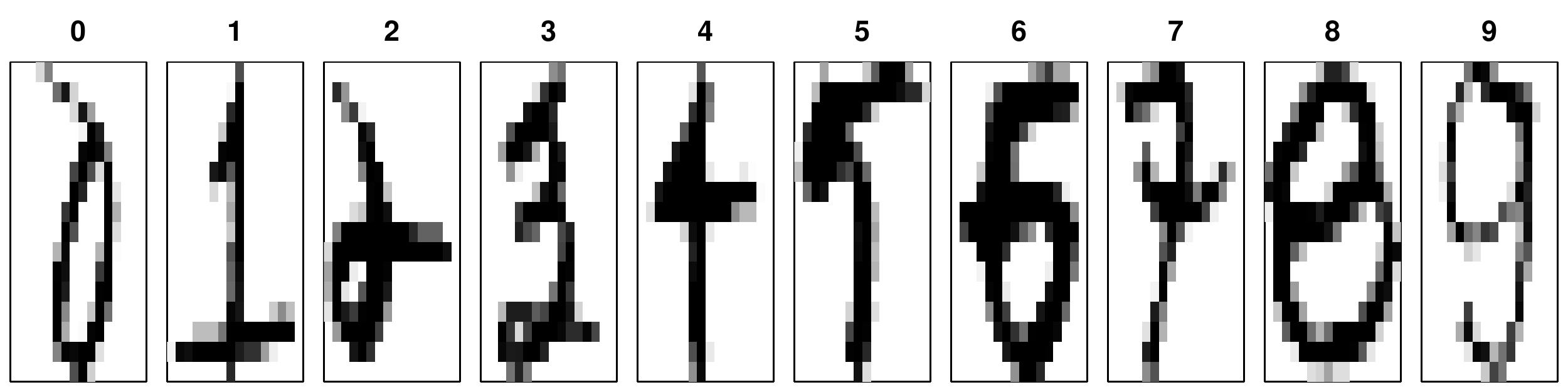}
\caption{Most outlying digits of each group.}
\label{fig:digout}
\end{figure}

This ``typical'' shape of the handwritten digits mentioned above 
can be extracted from our results. According to Equation~(\ref{eq:MultResRD})
we can compute for every observation of the training data a robust 
Mahalanobis distance. Figure~\ref{fig:digin} shows the images with the 
smallest distances per class, and thus these would correspond to the most
``inlying'' and thus typical writing styles of the digits. On the other hand, Figure~\ref{fig:digout}
shows the most outlying training set observations per class, which are those
observations of each class with the biggest distance.

\begin{figure}[htp]
\includegraphics[width=\textwidth]{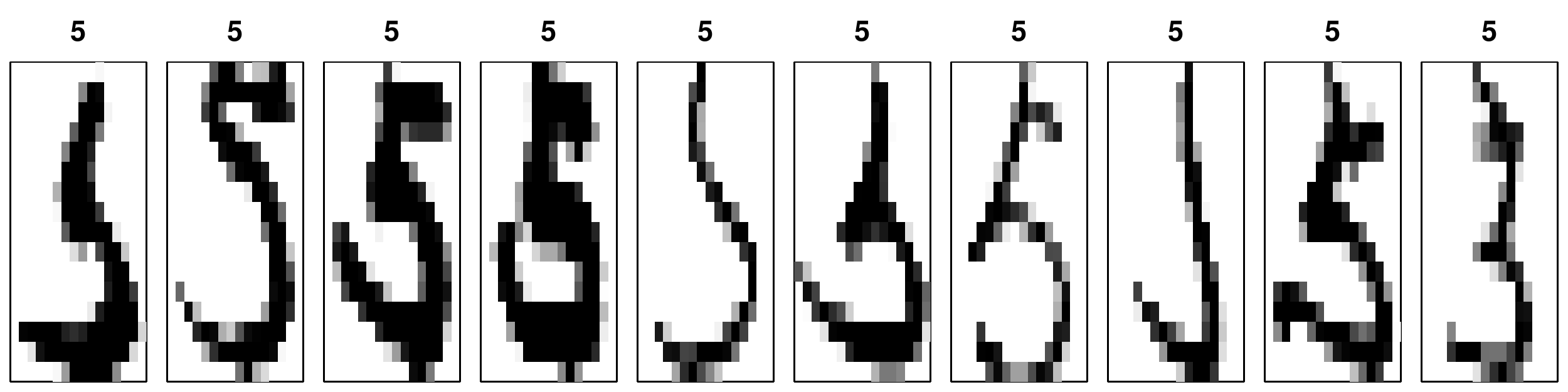}
\caption{Digits ``5'' from the test set, wrongly classified as ``3''.}
\label{fig:wrong53}
\end{figure}

Figure~\ref{fig:wrong53} shows the ten observations from the test set which should represent 
digits ``5'', but are incorrectly classified as digit ``3'', see also Table~\ref{tab:digittest}.
Looking at the coefficients in Figure~\ref{fig:coef}, the lower part in the image is very
similar, but the upper part should indicate the major differences; in particular, digit ``5''
is expected to have dark pixel information in the upper right corner, which is not the case
for digit ``3''. Indeed, this is what most of the images in Figure~\ref{fig:wrong53} are
missing. Other digits show different major differences to an expected ``5'' also mainly
in the upper part.

\subsection{Analysis of the fruit data set}

This data set has been used previously in the context of robust discrimination, for example
in~\citep{hubert2004fast}. It contains spectral information with 256 wavelengths,
thus is high-dimensional, for observations from 3 different cultivars of the 
same fruit, named D, M, and HA, with group sizes 490, 106, and 499. 
Group D in fact consists of two sub-groups, because for 140 observations
a new lamp has been installed in the measurement device. As we have no information
about the membership of these subgroups, we treat them as one group. Also
group HA should consist of 3 sub-groups due to a change of the illumination
system, and also this group was treated as a single group.

\begin{figure}[htp]
\includegraphics[width=\textwidth]{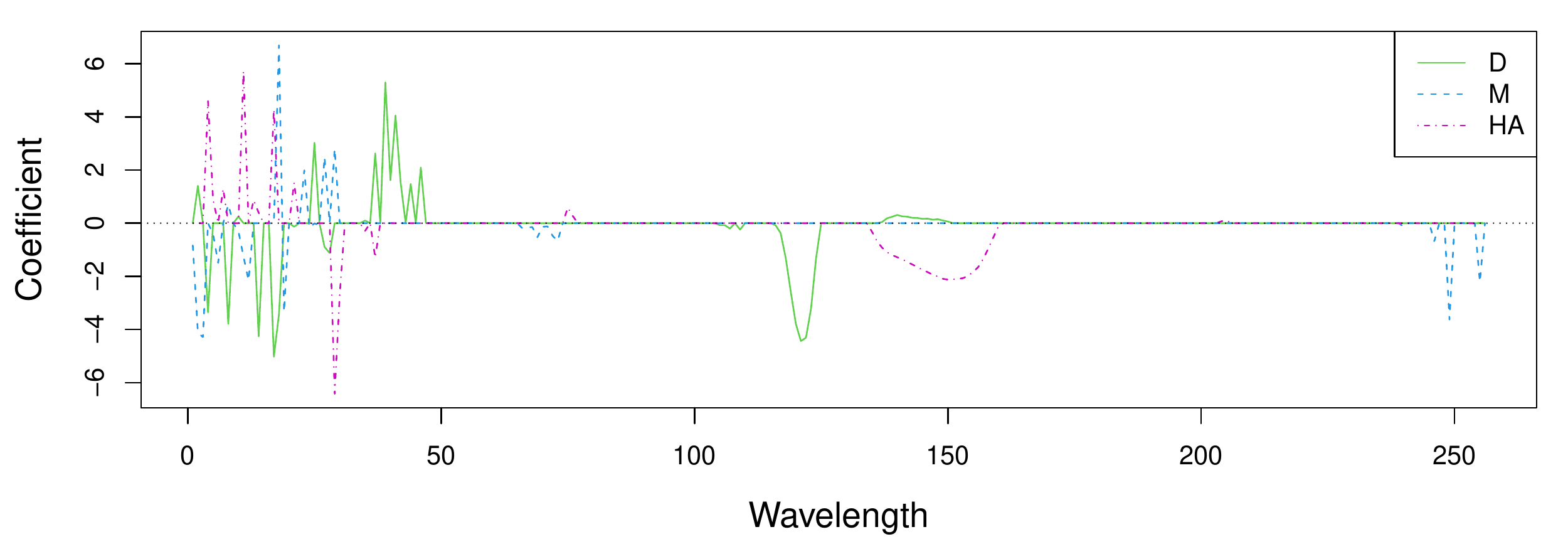}
\caption{Estimated coefficients for the fruit data set.}
\label{fig:fruitcoef}
\end{figure}

Applying our estimator yields the estimated coefficients visualized in
Figure~\ref{fig:fruitcoef}, which shows the 256 values per group as
lines. One can see that the model yields a very sparse solution.
The first wavelength range which looks rather confusing seems
to be important for the classification task.

The model has a correct classification rate of about 87\%, well balanced
over the groups. However, we did not go into a deeper evaluation of the
classification performance by using cross-validation.
Five-fold cross-validation has been used inside the procedure
for selecting the optimal tuning parameters. Here we have forced $\alpha$
to be higher, yielding more sparsity. With the optimal value of $\alpha$ we
would have had no sparsity at all, but an improved error rate of about
95\%.

\begin{figure}[htp]
\includegraphics[width=\textwidth]{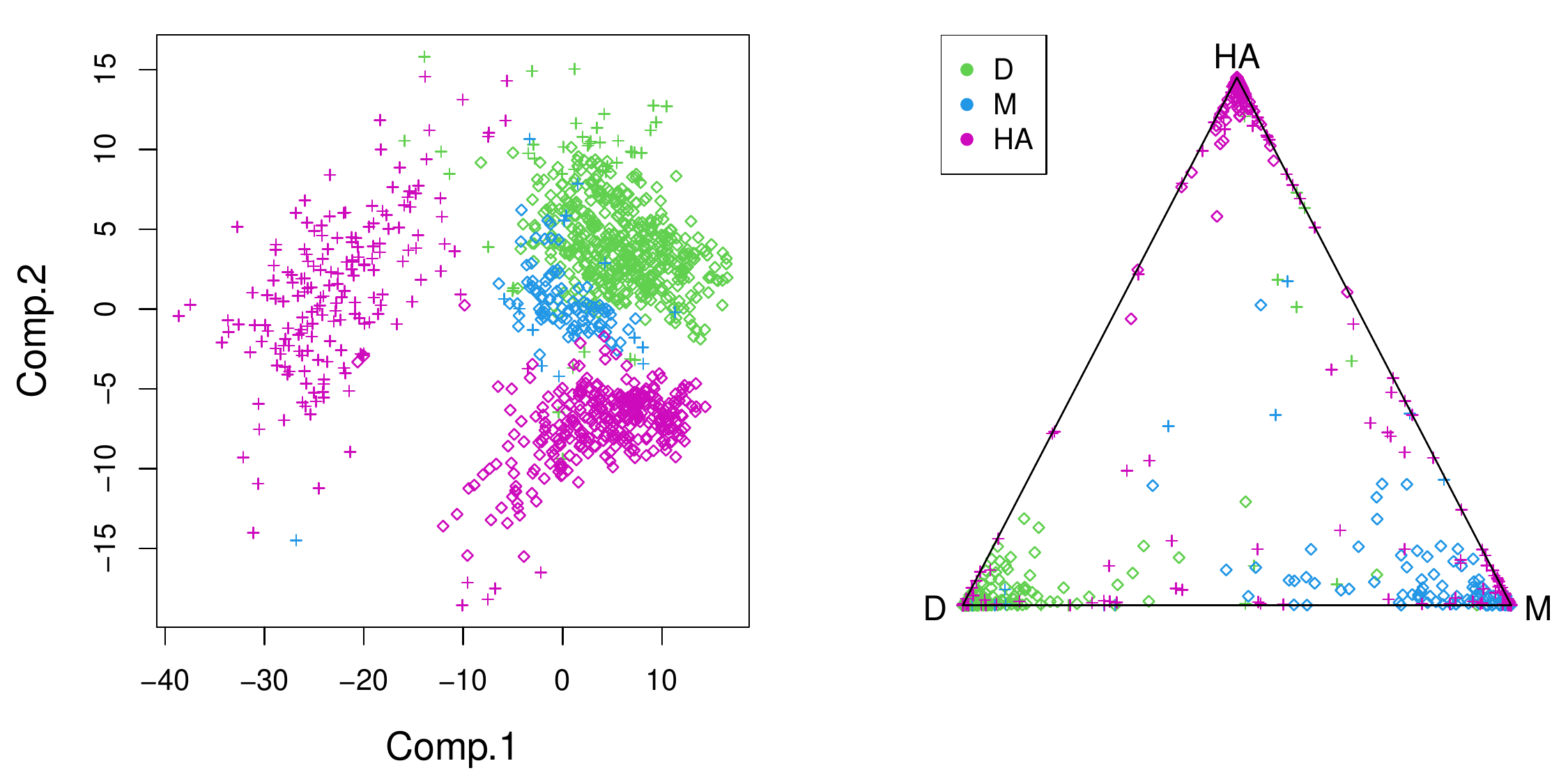}
\caption{First two principal components of the estimated scores (left)
and classification probabilities (right)
for the fruit data set.}
\label{fig:fruitscores}
\end{figure}

We computed the scores matrix $\mathbf{Z}_l$, for $l=1,2,3$,
see Section~\ref{sec:csteps}, and the resulting scaled robust distances,
see Equation~(\ref{eq:MultResRD}). Figure~\ref{fig:fruitscores} (left)
shows the scores of all groups in the space of the first two 
principal components, explaining nearly all of the variability.
The symbol colors are according to the group memberships, and the
symbols according to the scaled distance: a rhombus if this distance was 1,
and a ``+'' otherwise. For group HA we can indeed see that a big
part of the observations has been identified as outliers,
or at least as unusual observations. 
Also for group D we can see several outliers.
The right plot of this figure
shows a ternary plot with the estimated probabilities, again using the 
same symbols as before. Most observations are close to the edge of the triangle,
and thus they are correctly assigned to the corresponding class. 
Many of the outliers from the HA group are wrongly assigned to group M or D,
which also reduces the correct classification rate. If one would exclude the observations
with scaled distance smaller than one from the error rate calculation, the
correct classification rate would be higher than 99\%.

\section{Conclusions} 
\label{sec:conclude}

This paper introduced a robust and sparse method for multinomial regression. The method is able to identify outliers in
high-dimensional data, as well as mislabeled observations, thus observations
which rather belong to a different group, and it downweights such 
outliers in the estimation procedure. In contrast to linear discriminant
analysis, the multinomial regression model directly specifies the 
parameters which relate the variables to the classification problem, and
this supports the interpretability of the estimated parameters. 
In a sparse setting, non-zero parameters will be connected to informative variables, while uninformative noise variables are associated with
zero parameters. In presence of outliers, a robust method is supposed to not only return reliable
parameter estimates, but also a reliable identification of 
relevant and noise variables.

The idea to achieve robustness for the parameter estimation is 
based on trimming the penalized negative log-likelihood 
function~\citep{Friedman10}, similar as it has been proposed for robust
logistic regression in high dimensions~\citep{Kurnaz18}. 
Outliers are identified in the space of the scores, which are the values of the linear link function. The score space has at most dimension $K-1$,
where $K$ is the number of groups, and thus group-wise robustly estimated (and scaled)
Mahalanobis distances can be used for the purpose of outlier identification.
The score space is also very useful for visual data exploration and 
interpretation.
Finally, the outlyingness information can be incorporated in weights to
obtain a reweighted estimator which achieves higher efficiency than the 
trimmed estimator.
Sparsity is obtained by using an elastic net penalty, which results in an
intrinsic variable selection property besides dealing with 
the multicollinearity problem, and therefore the proposed method is very useful in high-dimensional sparse settings. 

We have conducted simulations to compare the proposed estimator with its
non-robust counterpart introduced in~\cite{Friedman10}. Hereby, various
scenarios such as $n>p$, $p>n$, and increasing sparsity levels have
been considered. For uncontaminated data, the robust estimator 
sometimes tends to lead to a higher false negative rate than the 
classical estimator, and thus to a sparser model. In contrast, the false
positive rate is often smaller than for the classical estimator,
which means to obtain fewer false discoveries. Again, depending on the
setting, the robust estimator can lead to a (slightly) increased 
misclassification error. In presence of contamination we have seen that
the robust estimator leads to a performance which is very similar to the
uncontaminated case, while the classical estimator is severely influenced
by the outliers.

Finally, the real data applications in Sec.~\ref{sec:applications} 
revealed the usefulness of the proposed estimator. Plots of the estimated
regression parameters lead to very interesting conclusions, since we can see which variables (pixels, wavelengths)
are important for the classification task. Also the outlyingness
information and the visualization of the scores is highly useful for 
a deeper understanding of the problem.

The algorithm for computing the estimator has been implemented in the R package \textit{enetLTS}~\citep{Kurnaz18R}. 
This package is using internally the R package \textit{glmnet} \citep{Friedman21R} which also implements Poisson, Cox and multivariate regression. As a matter of course, in our future work 
we plan to extend the introduced algorithm to these models. 

\section*{Acknowledgments} 

This work is supported by grant TUBITAK 2219 from the Scientific and Technological Research Council of Turkey (TUBITAK).

\end{document}